\documentstyle{article}

\newcommand{\pl}{\partial}
\newcommand{\bq}{\begin{equation}}
\newcommand{\eq}{\end{equation}}
\newcommand{\vb}{{\bf v}}
\newcommand{\rb}{{\bf r}}
\begin{document}

\centerline{\large \bf Microscopic discontinuity of fluids}
\vskip 0.2in
\centerline{Dept. of Physics, Beijing University of Aeronautics}
\centerline{and Astronautics, Beijing 100083, PRC}
\centerline{C.Y. Chen, Email: cychen@public2.east.net.cn}

\vskip 0.2in

\noindent {\bf Abstract:}
We reveal that realistic fluids generate microscopic-level discontinuity constantly and the discontinuity spreads out with motion of particles rather rapidly and widely. These things cannot be treated by the standard kinetic equations, and thus the existing fluid theories, macroscopic ones and microscopic ones, need to be revised considerably.

\vskip 0.1in
\vskip 0.5in

It has been unanimously believed that the following standard  framework\cite{harris}\cite{ichimaru}
\vskip 0 pt\hskip 15pt Continuity equations and Liouville's theorem 
\vskip 0 pt\hskip 35pt $ \downarrow$
\vskip 0 pt\hskip 15pt BBGKY hierarchy equations
\vskip 0 pt\hskip 35pt $\downarrow$
\vskip 0 pt\hskip 15pt Kinetic equations $\rightarrow$ Fluid equations
\vskip 0 pt\hskip 35pt $\downarrow\hskip 80pt \downarrow$ 
\vskip 0 pt\hskip 15pt Equilibrium and nonequilibrium phenomena 
\newline
gives sound microscopic and macroscopic descriptions of all classical fluids observed in the physical reality. 

Nevertheless, discoveries in various fields, such as those related to turbulence, chaos and dissipative structures, constantly revealed fluid phenomena that appear to be inconsistent with our existing theoretical knowledge. In the studies of statistical behavior of charged particles\cite{chen1}\cite{chen2}, we ourselves were in trouble to construct intuition out of what was obtained from the kinetic equations. The situation motivated us to make a lot of investigation on related fundamental subjects and finally  resulted in our critical attitude toward the existing kinetic theory.

In the usual textbook elaboration of the kinetic theory\cite{reif}\cite{harris}, all the particles are at first assumed to make motion in the way as if they constitute a continuous medium and the mediumlike behavior is cast into a differential operator similar to that in fluid mechanics; an integral operator, as a correction term, is then attached, in which particles interact with each other as if they are ordinary colliding particles. In our view, the formalism, though seems to give a comprehensive description of the two contradictory aspects, medium aspect and particle aspect, runs into conceptual difficulties. According to the theory, if the number of particles becomes more and more (the gas becomes denser and denser), then the collisional term becomes overwhelming; otherwise, the medium aspect becomes more important. (In some textbook, it is explicitly stated that if the collisions between particles are not significant the Boltzmann equation is reduced to the collisionless Boltzmann equation, called sometimes the Vlasov equation.) However, this picture is not in harmony with the simple physical intuition which says that a ``fluid'' consisting of a very small number of particles behaves as a collection of particles only. The particles in such a ``fluid'' move like individual particles and collide with each other like individual particles; no mediumlike character can be observed.

In this letter, we will investigate the issue thoroughly. It will be shown that a realistic Boltzmann gas (by ``Boltzmann gas'' we mean a gas in which interforces between particles are short-ranged) is capable of generating several kinds of discontinuous distribution functions and the dynamical behavior of such distribution functions is far from mediumlike and cannot be described by the existing kinetic theory.

We start our discussion by recalling the standard kinetic equations briefly. These equations can be written in the following unifying form
\bq \label{eq0}\left(\frac{\pl}{\pl t}+\vb \cdot\frac{\pl}{\pl \rb}+\frac{{\bf F}}m\cdot \frac {\pl}{\pl \vb}\right) f=\left(\frac{\delta f}{\delta t}\right)_{\rm collision},\eq
where the right side represents an integral collisional operator, which takes different forms for different kinetic equations. For purposes of this paper, we confine ourselves to a gas in which collisions are not significant. Under this understanding, Eq. (\ref{eq0}) becomes      
\bq \label{eq1} \left(\frac{\pl}{\pl t}+\vb \cdot\frac{\pl}{\pl \rb}+\frac{{\bf F}}m\cdot \frac {\pl}{\pl \vb}\right) f= 0.\eq 
Moving together with one particle in the fluid, we simply find that the equation can be expressed as a convective derivative 
\bq\label{inv} \left( \frac{df}{dt}\right)_{\rm path}=0.\eq
In words, this is to say that the distribution function keeps invariant along any particle path in the six-dimensional phase space ($\mu$-space).

The invariance expressed by (\ref{inv}) is quite amazing in the sense that the distribution function can have such nice and regular behavior while each particle in the fluid seems free to make any kinds of motions, including some ``irregular ones''. This is by no means possible and seeking for hidden requirements is kind of necessary. A careful inspection leads us to the following three requirements. (i) Throughout the dynamical process, the distribution function must be continuous both in terms of position and in terms of velocity. (ii) The force term ${\bf F}$ in the equation must be free from dissipation. (iii) No stochastic forces (collision-type) get involved, namely the force ${\bf F}$ must be sufficiently smooth in space. In what follows, we are mainly concerned with the first requirement and it will be shown that a realistic fluid can constantly generate discontinuity at the miroscopic level and the discontinuity spread out rather rapidly and widely.

\setlength{\unitlength}{0.01in} 
\begin{picture}(200,185)
\put(15,155){\makebox(35,8)[l]{\bf Figure 1}}
\put(88,15){\makebox(35,8)[l]{\bf (a)}}
\put(268,15){\makebox(35,8)[l]{\bf (b)}}
\put(98,90){\line(0,1){40}}
\multiput(98,60)(0,10){6}{\line(2,1){30}}
\multiput(98,80)(0,-10){3}{\line(-2,-1){30}}
\multiput(118,70)(0,10){6}{\vector(-2,-1){5}}
\put(278,90){\line(0,1){40}}
\multiput(278,60)(0,10){6}{\line(2,-1){30}}
\multiput(278,80)(0,-10){3}{\line(-2,1){30}}
\multiput(298,50)(0,10){6}{\vector(-2,1){5}}
\end{picture}

Firstly, consider a piece of boundary shown in Fig. 1. On the right side of it, there are particles belonging to an ordinary gas initially. Suppose that we are able to observe these particles continuously. It is easy to find that the distribution function at every point on the left side of the boundary involves discontinuity in terms of velocity angle. 

\setlength{\unitlength}{0.01in} 
\begin{picture}(200,165)
\put(15,135){\makebox(35,8)[l]{\bf Figure 2}}
\multiput(90,15)(10,0){2}{\line(0,1){50}}
\multiput(90,70)(10,0){2}{\line(0,1){45}}
\multiput(90,40)(0,5){12}{\line(1,0){10}}
\put(105,70){\vector(2,1){40}}
\put(105,67.5){\vector(2,0){40}}
\put(105,65){\vector(2,-1){40}}
\end{picture}

Then, let's look at the particles leaking through a small hole of a gas container (free-expansion gas), shown in Fig. 2. We can find that the distribution function related to all these particles is continuous in the position space but not in the velocity space. 

\setlength{\unitlength}{0.01in} 
\begin{picture}(200,120)
\put(15,95){\makebox(35,8)[l]{\bf Figure 3}}
\put(70,20){\line(1,0){60}}
\multiput(70,20)(5,0){13}{\line(0,-1){5}}
\multiput(95,20)(10,0){2}{\line(1,1){50}}
\multiput(95,20)(10,0){2}{\line(-1,1){50}}
\multiput(95,20)(10,0){2}{\vector(1,1){25}}
\multiput(75,40)(10,0){2}{\vector(1,-1){5}}
\end{picture}

Actually, there are more sources of discontinuity. To get a complete knowledge about this, it is instructive to study another kind of interaction between particles and boundaries, collisions between particles and boundaries. Such collisions are usually assumed to be kind of deterministic and the assumption is adopted by many college-level textbooks\cite{orear}, in which molecules are considered as having hard-spheres, walls as perfectly smooth and collisions as classically elastic, shown in Fig. 3. According to this elastic model, the distribution function must keep invariant after the collisions, which partly explains why boundary problems have not received much attention in the standard kinetic theory.  

\setlength{\unitlength}{0.01in} 
\begin{picture}(200,160)
\put(15,135){\makebox(35,8)[l]{\bf Figure 4}}
\put(50,20){\line(1,0){100}}
\put(100,20){\line(0,1){80}}
\multiput(50,20)(5,0){21}{\line(0,-1){5}}
\put(50,60){\vector(1,0){20}}
\put(100,30){\vector(1,0){5}}
\put(100,40){\vector(1,0){10}}
\put(100,50){\vector(1,0){15}}
\put(100,60){\vector(1,0){17}}
\put(100,70){\vector(1,0){18.5}}
\put(100,80){\vector(1,0){19.3}}
\put(100,90){\vector(1,0){20}}
\end{picture}

It is easy to see that the model outlined above is an ideal one serving for educational purposes. Actual observations in fluid mechanics tell us a  different story\cite{fung}: the fluid particles immediately next to a solid surface remain almost stationary with respect to the surface (apparently in the average sense), shown in Fig. 4. This is called the no-slip condition by scientists in fluid mechanics.

\setlength{\unitlength}{0.01in} 
\begin{picture}(200,120)
\put(15,95){\makebox(35,8)[l]{\bf Figure 5}}
\put(70,20){\line(1,0){60}}
\multiput(70,20)(5,0){13}{\line(0,-1){5}}
\put(95,20){\vector(2,1){40}}
\put(95,20){\vector(1,1){30}}
\put(95,20){\vector(1,2){20}}
\put(95,20){\vector(0,1){50}}
\put(95,20){\vector(-2,1){40}}
\put(95,20){\vector(-1,1){30}}
\put(95,20){\vector(-1,2){20}}
\end{picture}

To conform to the no-slip condition, a different model, called the statistical model, should be adopted. In this model, particles leave, after collisions, the solid surface in the way as if they are emitted from there, and their velocities will distribute according to a statistical law irrespective of how they come to there.
(Though more sophisticated models may be better off in the academic sense, we take this simple one as our ``zeroth-order approximation'' for relative simplicity.) 

Taking this statistical model implies that we are ready not only to admit that interaction between particles and boundaries is of a dissipative nature and stochastic nature but also to admit that the distribution functions of emitted particles are discontinuous: if the area of the surface is small enough the situation is like the gas leaking out of a container in Fig. 2; if the surface is of a finite-size, the distribution function will be roughly the same as that created by the finite boundary in Fig. 1. 

Examples aforementioned have shown that realistic boundaries can indeed generate, by blocking particles or by emitting particles, certain kinds of discontinuous distribution functions. A conception which may come to one's mind immediately is that by introducing the $\delta$-function and/or the step function these discontinuous distribution functions may be treatable  as if they are continuous ones. Such conceptions got a lot of successes in other physical fields, why not in this field? We will see, however, that the position-velocity phase space, in which distribution functions are defined, possesses special features and one of the features is that the position $\rb$ and the velocity $\vb$ are assumed to be completely independent of each other. (Note that we have $\vb=\dot \rb$ for each particle in the fluid.)  It is because of this assumption that the existing kinetic theory is inapplicable in the situations even if the $\delta$-function and step-function are properly introduced.  

We investigate the following distribution function
\bq\label{delta} f=\frac{n_0}{r^2}\delta(\vb-v_0\frac{\rb}r ),\eq
where $n_0$ and $v_0$ represent two constants, $\rb$ the position vector and $r=|\rb|$. By referring to Fig. 2, we can recognize that the distribution function characterizes a kind of free expansion gas. The time development of these particles can be expressed by
\bq \begin{array}{l} x^\prime= x+v_x t\\y^\prime= y+v_y t\\z^\prime= z+v_z t \end{array} \eq
and 
\bq  v_x^\prime= v_x,\quad v_y^\prime= v_y, \quad v_z^\prime=v_z. \eq
By a direct calculation, we know that along a particle trajectory 
\bq \frac {f(t,\rb,\vb)}{f(t^\prime, \rb^\prime,\vb^\prime)}=\frac {{r^\prime}^2}{r^2} .\eq
Actually, Fig. 2 can directly tell us that, in contrast with (\ref{inv}), the distribution function is not invariant along the particle trajectories.

The difficulty becomes even more striking if we substitute the distribution function (\ref{delta}) into the kinetic equation (\ref{eq1}). The following two partial derivatives
\bq\left.\frac{\pl[\delta(\vb-v_0\rb/r)]}{\pl \rb}\right|_\vb =? \eq
and
\bq\left.\frac{\pl[\delta(\vb-v_0\rb/r)]}{\pl \vb}\right|_\rb =? \eq
are rather puzzling. It is obvious that the troublesome situation arises not because we do not know how to differentiate the $\delta$-function but because we cannot determine whether or not $\rb$ and $\vb$ correlate.

Up to now, we have seen that such discontinuous distribution functions pose difficulty in the two related respects: (i) The discontinuity spreads throughout almost the entire space instead of ``staying'' on certain boundaries. (ii) Since the discontinuity spreads the distribution function must contain path information of particles; this is in conflict with the partial-differential equations where $\rb$ and $\vb$ are considered independent of each other. The latter respect, called the paradox of $\rb$-$\vb$ correlation in this letter, holds its significance rather generally.

Before finishing this letter, it is important and relevant to look at whether or not other related theories, such as the continuity equations and Liouville's theorem, hold in the presence of such  discontinuity. (In fact, by referring to the paradox of $\rb$-$\vb$ correlation, readers themselve can come up with the answer to the question without many studies.)   
Adopting that the force term in (\ref{eq1}) is independent of velocity, we find that the collisionless Boltzmann equation (\ref{eq1}) is identical to the continuity equation in $\mu$-space
\bq \label{eq2} \frac{\pl f}{\pl t}+\frac{\pl (\vb f)}{\pl\rb}+ \frac {\pl (\dot \vb f)}{\pl \vb}= 0.\eq 
It is then obvious that the continuity equation in $\mu$-space does not generally hold for discontinuous distribution functions.

In a similar way, it can be shown that the continuity equation in the grand phase space ($\Gamma$-space), which can be regarded as a natural extension of the continuity equation in $\mu$-space, is not valid for systems having discontinuous distribution functions.     

Finally, we study a more fundamental issue: Is Liouville's theorem applicable in statistical mechanics? To get a definite answer to the question and, at the same time, to gain insight into the important subject, it is necessary to examine derivations of the theorem in detail.

In textbooks, one of the derivations is based on the continuity equation in $\Gamma$-space and will be discussed no more. Another derivation\cite{landau}, which invokes the Hamiltonian formalism and the Jacobian approach, is worth our special attention.

Suppose that the Hamiltonian of a system with $2N$ degrees of freedom is in the form
\bq \label{hamil}
H=\sum\limits_{N} \frac{p^2_i}{2m}+\Phi(q_1,q_2,\cdots,t).\eq
We have explicitly assumed that the potential term $\Phi$ only contains the position coordinates $q_1,q_2,\cdots$ and implicitly assumed that the potential term $\Phi$ is differentiable in terms of these coordinates. Note that these two things have defined the syatem as a pure mechanical one with no dissipative and stochastic forces involved. Suppose that $d\Gamma$ represents a phase volume element in $\Gamma$-space. One version of Liouville's theorem states that the value of $d\Gamma$ is motion-invariant.

To prove it, we adopt that the time-development of the system is a canonical transformation 
\bq p(t_0),q(t_0) \rightarrow P(t),Q(t).\eq
In the grand space we have, generally,
\bq\label{volume} dqdp=DdQdP,\eq
where 
\bq D=\left\| \frac{\pl (Q,P)}{\pl (q,p)} \right\|\eq
is the Jacobian of the transformation. By applying the properties of the canonical transformation, we get (with the details omitted)
\bq D=1 . \eq
Equation (\ref{volume}) can then be written in the form 
\bq\label{deq} d\Gamma_0(t_0) =d \Gamma(t).\eq

Eq. (\ref{deq}) is regarded as Liouville's theorem in classical mechanics. To have Liouville's theorem in statistical mechanics, which states that the grand density $\rho$ conserves in statistical processes, several requirements have to get involved. By no coincidence, the requirements are almost exactly the same as those we listed for the invariance of distribution functions: In the entire statistical process the grand density $\rho$ has to keep continuous and the external forces, which are associated with $\Phi$ in (\ref{hamil}), must be completely free from the dissipative nature and stochastic nature. As has been shown, these requirements cannot be generally fulfilled for realistic statistical systems.
All these tell us that Liouville's theorem should be regarded as a theory in classical mechanics only.

In summary, this letter has revealed that seemingly simple fluids may actually be beyond the scope of the existing fluid mechanics and beyond the scope of the existing kinetic theory. 

Discussion with Professor Keying Guan is gratefully acknowledged. His mathematical viewpoint on turbulence is a stimulating factor of this letter. This work is partly supported by the fund provided by Education Ministry, PRC.

\end{document}